\documentstyle[aps,prb,psfig]{revtex}

\begin{document}
\draft
\twocolumn[\hsize\textwidth\columnwidth\hsize\csname @twocolumnfalse\endcsname
\title{Twin Boundaries in $d$-wave Superconductors}
\author{D.L.~Feder, A.~Beardsall, A.J.~Berlinsky, and C.~Kallin}
\address{Department of Physics and Astronomy, McMaster University, Hamilton,
Ontario L8S 4M1, Canada}
\date{\today}
\maketitle

\begin{abstract}

Twin boundaries in orthorhombic $d$-wave superconductors are investigated
numerically using the Bogoliubov-deGennes formalism within the context of an 
extended Hubbard
model. The twin boundaries are represented by tetragonal regions of variable 
width, with a reduced chemical potential. For sufficiently large twin boundary
width and change in chemical potential, an induced $s$-wave component may break 
time-reversal 
symmetry at a low temperature $T^*$. The temperature $T^*$, and the magnitude
of the complex component, are found to depend strongly on electron density. 
The results are compared with recent tunneling measurements.

\end{abstract}
]

In spite of mounting experimental evidence that the high-temperature
superconductors have an order parameter with $d_{x^2-y^2}$ 
($d$-wave) symmetry,\cite{revs} a number of experiments on both twinned and
untwinned YBa$_2$Cu$_3$O$_{7-\delta}$ (YBCO) suggest the 
presence of an additional $s$-wave order parameter.\cite{chaud,dynes} By 
symmetry, a small $s$-wave component always coexists with a 
predominantly $d$-wave order parameter in an orthorhombic superconductor such as
YBCO, and changes its sign (relative to the $d$-wave component) across a twin 
boundary.\cite{walker}
The experimental results can be understood either if the $s$-wave component 
of the order parameter breaks time-reversal ($\cal T$) symmetry near the twin 
boundary at low temperatures, as it can near 
surfaces,\cite{kirtley,greene,sigrist1,sigrist2,matsumoto,sauls}
or if there is far more of one kind of twin domain (i.e.\ twin
boundaries form in groups).\cite{rosova,gaulin} Recent SQUID 
measurements on vortices trapped by twin boundaries in YBCO
did not detect the fractional flux that would accompany local 
$\cal T$-violation.\cite{moler} In the present work, however, we present 
strong evidence that such a symmetry breaking could indeed occur in the 
vicinity of twin boundaries at low temperatures under certain conditions.

Sigrist {\it et al.}\cite{sigrist1} have addressed the possibility of 
${\cal T}$-violation near boundaries by considering the Ginzburg-Landau
(GL) free energy for a homogeneous orthorhombic $d$-wave superconductor:

\begin{eqnarray}
F_s&=&F_n+\alpha_d|d|^2+\alpha_s|s|^2+\beta_1|d|^4+\beta_2|s|^4+\beta_3|s|^2
|d|^2\nonumber \\
&+&\beta_4({s^*}^2d^2+s^2{d^*}^2)+\beta_5(s^*d+sd^*),
\end{eqnarray}

\noindent where $s$ and $d$ are the $s$-wave and $d$-wave components of a
superconducting order parameter of the form $d+e^{i\theta}s$, with $\theta$
the relative phase between $s$ and $d$. Only the lowest-order orthorhombic term
is kept in the GL free energy since $s\ll d$ for small $\beta_5$. Assuming all
the coefficients are positive (except $\alpha_d$), the $\beta_5$ term favors 
$\theta=n\pi$ ($n$ integer), while the $\beta_4$ term favors $\theta=\pi/2$. 
Thus, in a tetragonal superconductor there can be 
a continuous transition to a bulk ${\cal T}$-violating phase at a 
temperature $T^*$ given by $\alpha_s(T^*)=\left[2\beta_4(T^*)
-\beta_3(T^*)\right]d^2(T^*)$.\cite{sigrist2}
The coefficients of the GL free energy can be chosen such that 
$T^*\leq 0$ for a uniform system; the orthorhombic term further discourages a
positive $T^*$. Yet the suppression of $d$ near an inhomogeneity, or the 
presence of a tetragonal region within an orthorhombic system, could induce
local ${\cal T}$-violation for a finite temperature $0\leq T\leq T^*$. 
Furthermore, the associated complex $s$-wave component could be large relative
to $|d|$, and would vary on a new length scale.\cite{sigrist1} In contrast, an 
$s$-wave component nucleated solely through spatial variations of $d$, such as
is found in magnetic vortices\cite{franz1,ren} or near impurities,\cite{franz2}
is usually small relative to $|d|$ (unless $\alpha_s\rightarrow 0$, which only 
occurs for densities just above a crossover to bulk $s$-wave 
superconductivity\cite{feder}) and varies on a length scale of the $d$-wave 
coherence length.

It is not clear, however, whether GL theory (which is strictly valid only
near the bulk superconducting transition temperature) can reliably describe the
low-temperature regime associated with ${\cal T}$-violation. In the present 
work, twin boundaries in $d$-wave superconductors are investigated numerically 
using
Bogoliubov-deGennes theory. We employ an extended Hubbard model which
gives rise to $d$-wave superconductivity in a restricted parameter 
regime.\cite{micnas}
Despite its simplicity, results obtained previously using 
this model\cite{pekka,wang} are
consistent with those obtained within the context of a model better 
representing the high-$T_c$ oxides,\cite{feder,dagotto} and with experimental 
results.\cite{keimer,maggio}
Twin boundaries are modeled as tetragonal 
regions of varying widths and reduced chemical potential, in order to 
approximate the experimental observations that twin boundaries are 
oxygen-deficient (i.e.\ locally antiferromagnetically insulating) regions, 
generally 7-40\AA wide.\cite{zhu}

\begin{figure}
\centerline{\psfig{figure=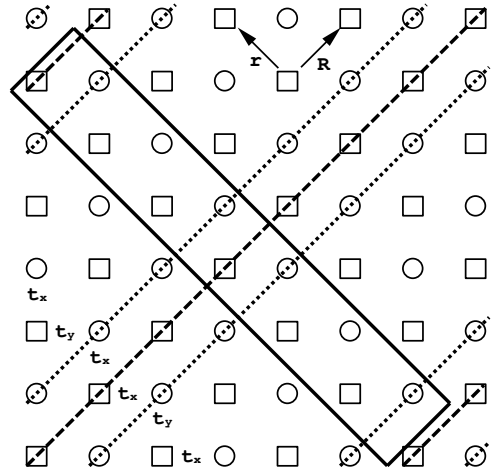,width=0.75\columnwidth}}
\caption{The unit cell of the finite-size system for the BdG calculations is 
shown as a solid 
line superimposed on a square lattice. Long and short dashed lines represent 
twin boundaries of width $0$ and $|\hat{r}|$, respectively. Basis points are 
labeled by circles and squares.}
\label{lattice}
\end{figure}

The Hamiltonian for the extended Hubbard model is:

\begin{eqnarray}
H&=&-\sum_{\langle ij\rangle\sigma}t_{ij}c^{\dag}_{i\sigma}
c^{\vphantom{\dag}}_{j\sigma}
-\mu\sum_{i\sigma}n_{i\sigma}
-\sum_{i\sigma}\mu^I_in_{i\sigma}\nonumber \\
&-&V_0\sum_in_{i\uparrow}n_{i\downarrow}-{V_1\over 2}
\sum_{\langle ij\rangle\sigma\sigma^{\prime}}
n_{i\sigma}n_{j\sigma^{\prime}},
\end{eqnarray}

\noindent where the sums are over spin and nearest-neighbors on the square 
lattice, $t_{ij}$ is a direction-dependent hopping parameter used to
model orthorhombicity, $\mu$ is the chemical potential, $\mu^I$ is a 
site-dependent impurity
potential representing the depletion of the carrier density at the twin 
boundary, and $V_0$ and $V_1$ are on-site
and nearest-neighbor interactions, respectively ($V>0$ denotes attraction).
Choosing the unit cell as shown in Fig.~\ref{lattice}, we can exploit the 
translational invariance of the Hamiltonian in the (110) direction.
With $\hat{R}\equiv\hat{x}+\hat{y}$ parallel and 
$\hat{r}\equiv -\hat{x}+\hat{y}$ perpendicular to the twin direction, we 
obtain the Bogoliubov-deGennes (BdG) equations:

\begin{equation}
\left(\matrix{\hat{\xi} & \hat{\Delta}\cr {\hat{\Delta}}^* & -\hat{\xi}\cr}
\right)\left(\matrix{u_{n,k}(r_{\alpha})\cr v_{n,-k}(r_{\alpha})\cr}\right)
=\varepsilon_{n,k}\left(\matrix{u_{n,k}(r_{\alpha})\cr v_{n,-k}(r_{\alpha})\cr}
\right),
\label{bdg}
\end{equation}

\noindent such that

\begin{eqnarray}
\hat{\xi}u_{n,k}(r_{\alpha})&=&-\sum_{\hat{\delta}}t_{\delta}u_{n,k}(r_{\alpha}
+\hat{\delta})\nonumber \\
&-&\left[\mu+\mu^I(r_{\alpha})\right]u_{n,k}(r_{\alpha}),
\end{eqnarray}

\begin{eqnarray}
\hat{\Delta}u_{n,k}(r_{\alpha})&=&\Delta_0(r_{\alpha})u_{n,k}
(r_{\alpha})\nonumber \\
&+&\sum_{\hat{\delta}}\Delta_{\delta}(r_{\alpha})u_{n,k}(r_{\alpha}
+\hat{\delta}),
\end{eqnarray}

\noindent where the gap functions are defined by

\begin{eqnarray}
\Delta_0(r_{\alpha})&\equiv&V_0\langle c_{\uparrow}(r_{\alpha})c_{\downarrow}
(r_{\alpha})\rangle;\\
\Delta_{\delta}(r_{\alpha})&\equiv&V_1\langle c_{\uparrow}(r_{\alpha}
+\hat{\delta})c_{\downarrow}(r_{\alpha})\rangle.
\end{eqnarray}

\noindent We have introduced the index $\alpha$ labeling the two basis points 
in the unit cell, the label $k$ which is the Fourier inverse of $R$, and
$\hat{\delta}=\hat{0},\hat{r},\hat{r}-\hat{R},-\hat{R}$ connecting nearest
neighbors with different basis indices. The equations (\ref{bdg}) are 
subject to the self-consistency requirements

\begin{equation}
\Delta_0(r_{\alpha})=V_0\sum_{nk}u_{n,k}(r_{\alpha})v^*_{n,-k}(r_{\alpha})
\hbox{tanh}\left({\beta\varepsilon_{n,k}\over 2}\right),
\end{equation}

\begin{eqnarray}
\Delta_{\delta}(r_{\alpha})&=&{V_1\over 2}\sum_{nk\hat{\delta}}\Big[u_{n,k}
(r_{\alpha}+\hat{\delta})v^*_{n,-k}(r_{\alpha})\nonumber \\
&+&u_{n,k}(r_{\alpha})v^*_{n,-k}(r_{\alpha}+\hat{\delta})\Big]
\hbox{tanh}\left({\beta\varepsilon_{n,k}\over 2}\right),
\end{eqnarray}

\noindent where the sum is over positive energy eigenvalues $\varepsilon_{n,k}$
only.

The orthorhombicity of YBCO is modeled by an anisotropy in the hopping 
parameters, reflecting the increased electronic mobility associated with the
chains;\cite{atkinson} throughout the present work we use $t_y/t_x=1.5$, which
approximates the observed $a$-$b$ anisotropy in the magnetic 
penetration depth,\cite{basov} and $V_0=-V_1=-3t_x$. Twin boundaries of up to 
$4|\hat{r}|$ width (corresponding 
to approximately 22\AA ) are investigated, with $t_y=t_x$ and $\mu^I\leq 0$
within the twin. The largest system studied is $100|\hat{r}|$ in length with 
400 $k$-states, or $80\,000$ sites; periodic boundary conditions are assumed 
throughout.

At a twin boundary of zero width and $\mu^I=0$, we find within the BdG 
theory that for all temperatures the dominant $d$-wave component of the order 
parameter is virtually unaffected. The extended and on-site $s$-wave components,
whose bulk values are approximately 10\% of $\Delta_d$, go from 
their near-bulk values to zero, over a single lattice spacing $r$, 
reversing their sign relative to $\Delta_d$ on either side of the boundary. As 
the impurity strength is increased at low temperatures, however, the $d$-wave 
and $s$-wave components become increasingly perturbed from their bulk 
values over the coherence length $\xi_{d+s}(T)$, where 
$\xi_{d+s}(0)\approx|\hat{r}|$ in the present work.
When the magnitude of the $d$-wave component in the twin boundary is suppressed 
to approximately half its bulk value, an additional complex $s$-wave component 
may be nucleated near the twin edge, breaking time-reversal symmetry. We have 
found no evidence for a phase transition to a bulk ${\cal T}$-violating state 
in a uniform system.

\begin{figure}
\centerline{\psfig{figure=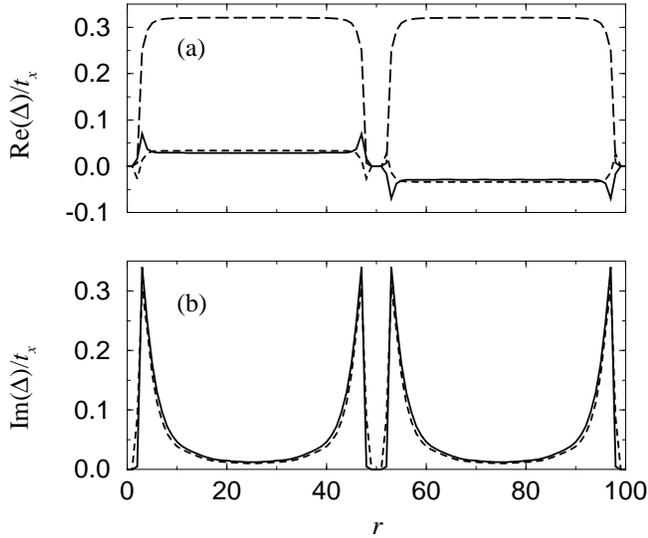,width=\columnwidth,angle=270}}
\caption{The (a) real and (b) complex components of the $d$-wave, on-site, and 
extended $s$-wave gap functions (corresponding to long-dashed, short-dashed, and
solid lines respectively) are shown as a function of distance $r$ perpendicular
to the twin boundary. In (b), ${\rm Im}(\Delta_s)$ is multiplied by a factor of
$-10$ to facilitate comparison with ${\rm Im}(\Delta_0)$. Results are obtained 
using $\mu=-t_x$, $\mu^I=-10t_x$, $T=0$, and twin width $4|\hat{r}|$. Twin 
boundaries are centered at positions $r=0$ (which is equivalent to $r=100$ by 
periodic boundary conditions) and $r=50$ in units of $|\hat{r}|$.}
\label{ops}
\end{figure}

The real and imaginary parts of the various components of the order parameter 
are shown in Fig.~\ref{ops} for a twin boundary with $\mu^I=-10t_x$, 
$\mu=-t_x$, $T=0$, and boundary width $W_T=4|\hat{r}|$. While all
the components go to zero rapidly within the twin boundary, both the
real and imaginary parts of the $s$-wave gap functions are enhanced near the twin 
edge. In the immediate vicinity of the twin boundary, the real $s$-wave 
components are perturbed from their bulk values over a short distance 
comparable to $\xi_{d+s}(0)$, reflecting the local nucleation of additional 
$s$-wave components through spatial variations of the dominant $d$-wave component. 
The presence of finite complex gap functions in the bulk, however, implies that 
the imaginary components vary over a different characteristic distance
$\xi_{is}(0)\gg\xi_{d+s}(0)$. This longer length scale, as well as the 
comparable sizes of ${\rm Im}(\Delta_0)$ near the twin edge and the bulk 
$\Delta_d$, is consistent with the GL predictions\cite{sigrist1} discussed 
above. 

\begin{figure}
\centerline{\psfig{figure=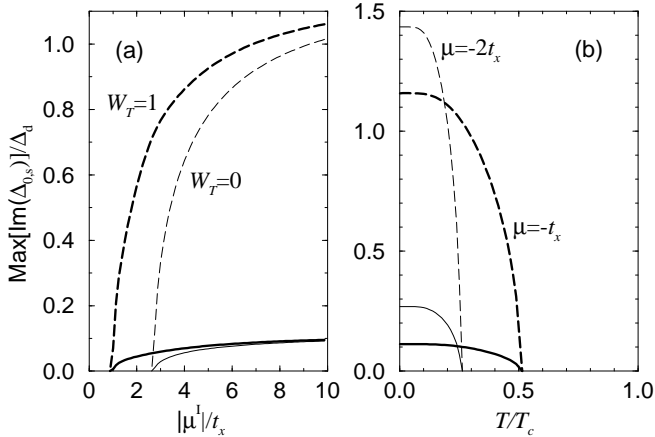,width=\columnwidth,angle=270}}
\caption{The maximum value of the complex on-site 
(dashed lines) and extended $s$-wave (solid lines) components relative to the 
bulk value of the $d$-wave order parameter are shown as a function of (a) 
$|\mu^I|$ (note $\mu^I\leq 0$) and $W_T$ at zero temperature and $\mu=-t_x$, 
and (b) $T/T_c$ and $\mu$. In (a) thin and bold
lines correspond to twin boundary widths $W_T/|\hat{r}|=0$ and 1, respectively. 
In (b), thin and bold lines denote results for $\mu=-2t_x$ and $\mu=-t_x$, and
we have chosen $\mu^I=-100t_x$, and $W_T=0$.}
\label{maxs}
\end{figure}

As shown in Fig.~\ref{maxs}, the size of the complex $s$-wave component 
nucleated near the twin boundary is extremely sensitive to the
temperature, impurity strength, and the width of the twin boundary. At zero
temperature, Fig.~3(a), a ${\cal T}$-violating state first appears for 
$\mu^I\approx -2.7t_x$ at all 
electron densities for an impurity line (i.e.\ $W_T=0$). As the
impurity strength continues to increase, the perturbation of the $d$-wave
component, and the maximum values of the imaginary $s$-wave components, begin to
saturate. For $W_T>0$, however, a lower impurity strength can give rise
to ${\cal T}$-violation at zero temperature, since the $d$-wave component is 
already suppressed by approximately $20\%$ in a locally tetragonal region 
(with $t_x=t_y$) even for $\mu^I=0$. Increasing $W_T$ beyond approximately
$3|\hat{r}|$ has no further effect. This result, valid for all 
electron densities, is also consistent with the GL prediction\cite{sigrist1} 
that local tetragonal symmetry could favor a time-reversal breaking state at 
low temperature. 

The growth of all the $s$-wave components with decreasing
chemical potential reflects the impending instability of the system against 
bulk dominant $s$-wave superconductivity at slightly lower electron 
densities.\cite{feder,micnas}
As the temperature is increased at finite $\mu^I$, the complex
component decreases to zero as $\sqrt{1-T/T^*}$; the transition temperature
$T^*$ is strongly density-dependent, scaling roughly with $\Delta_d$. The same
$T^*$ is obtained for wider twin boundaries at a given density, though the
magnitudes of the complex $s$-wave components increases with increasing $W_T$.

\begin{figure}
\centerline{\psfig{figure=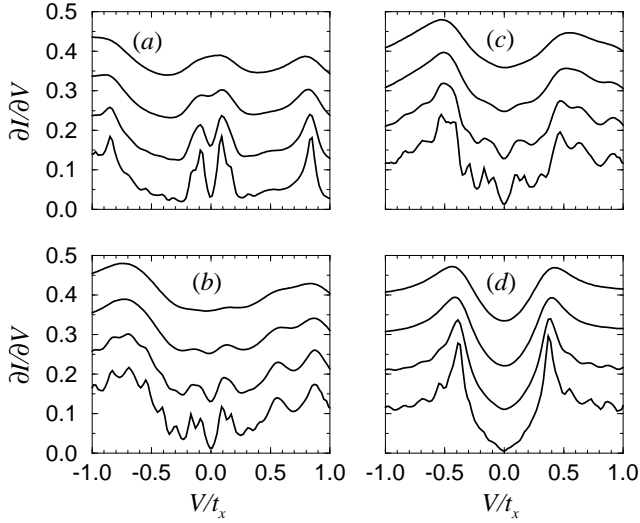,width=\columnwidth,angle=270}}
\caption{A low-energy portion of the tunneling conductance near a twin
boundary is shown as a function of energy, distance $r$ from the twin edge
where ({\it a}) through ({\it c}) correspond to $|\hat{r}|$ through
$3|\hat{r}|$ while $(d)$ illustrates the bulk, and temperatures $T=0$ (lower),
$T=0.05T_c$ (offset 0.1), $T=0.1T_c$ (offset 0.2) and $T=0.15T_c$ (offset 0.3).
Parameters are as in Fig.~\ref{ops}.}
\label{conduct}
\end{figure}

The spatial variation of the $s$-wave component's phase relative to $\Delta_d$
implied by Fig.~\ref{ops}
leads to currents which flow parallel to the twin surface and in opposite 
directions on either side of the twin boundary. The strong impurity potential
therefore mimics a line of temperature-dependent magnetic flux passing through
the twin boundary and oriented parallel to the $c$-axis. As shown by the 
differential conductance in Fig.~\ref{conduct}

\begin{eqnarray}
{\partial I(r)\over\partial V}&\propto&-\sum_{nk}\left[|u_{n,k}(r)|^2f'\left(V
-\varepsilon_{n,k}\right)\right.\nonumber \\
& &\qquad +\left.|v_{n,k}(r)|^2f'\left(V+\varepsilon_{n,k}\right)\right],
\label{tunn}
\end{eqnarray}

\noindent where $f'$ is the voltage-derivative of a Fermi function, this
effective field splits the low-energy band of virtual-bound states associated
with Andreev reflections at the twin surface.\cite{hu} Alternatively, the
presence of two low-energy quasiparticle peaks in the tunneling conductance at
low temperatures reflects the existence of a physical gap in the excitation
spectrum, proportional to the magnitude of the total complex $s$-wave 
component. The zero-temperature maximum peak-to-peak separation, found in
Fig.~\ref{conduct}a to be approximately $0.2t_x\sim 2\,{\rm meV}$ (where $t_x$
is estimated from $T_c=0.51t_x\approx 60\,{\rm K}$ at this carrier
concentration $\langle n\rangle\approx 0.75$), grows with
increasing $|V_0|$ and $|\mu^I|$ but diminishes with increasing temperature
and distance from the twin boundary. While the tunneling conductance
exhibits low-temperature features that are no doubt
finite-size effects, it is evident that the low energy band splits at a
temperature $0.1T_c\approx 6\,{\rm K}$ which is considerably
lower than the $T^*\sim 0.5T_c$ estimated from Fig.~\ref{maxs}b. A comparable
splitting of the zero-energy peak has been recently observed\cite{greene} in
tunneling spectra of YBCO surfaces, and has been interpreted\cite{sauls}
as a clear signature of ${\cal T}$-violation.

In summary, we have found evidence for time-reversal symmetry breaking near
twin boundaries in a $d$-wave orthorhombic superconductor at low 
temperatures $T<T^*<T_c$, where $T^*/T_c$ scales approximately with the size of 
the bulk $d$-wave gap. The magnitudes of the complex $s$-wave components associated 
with the ${\cal T}$-violation depend strongly on the chemical potential and
depletion of the carrier density in the twin boundary. These $s$-wave gap 
functions could
be responsible for the finite Josephson currents observed in c-axis tunnel 
junctions to heavily-twinned YBCO.\cite{dynes} As a consequence 
of the time-reversal breaking, the low energy quasiparticle peak in the
tunneling conductance (related to the zero-bias anomaly in STS) is predicted
to split in the vicinity of the twin edge.

\begin{acknowledgments}

The authors are grateful to D.~Branch, L.-W.~Chen, M.~Franz, L.H.~Greene,
K.A.~Moler, and M.I.~Salkola for their insightful comments. This work has been
partially supported by the Natural Sciences and Engineering Research Council
of Canada and the Ontario Centre for Materials Research.

\end{acknowledgments}

\end{document}